# A Sustainable Concept for Permafrost Thermal Stabilization

Egor Y. Loktionov[a,*], Elizaveta S. Sharaborova[a], Taissia V. Shepitko[b]

[a] Bauman Moscow State Technical University, 5-1, 2nd Baumanskaya str., Moscow 105005, Russia

[b] Russian University of Transport, 9b9, Obraztsova str., Moscow 127994, Russia

**Abstract**

Although permafrost thermal stabilization systems have been used for a long time already, they have always had shortcomings and limitations of performance which have become even more pronounced in the warming climate. Those could be overcome to some extent, but at high cost mainly defined by need in energy supply. We have suggested a novel concept combining improved energy efficiency and solar power to ensure significant reduction of the thawing layer (to 20 cm order). We have performed calculations for the broad range of permafrost conditions to compare traditional methods for thermal stabilization and their combination to the suggested concept. The latter only has ensured minimum thawing layer over summer even at the southern edge of permafrost extent. The importance of minimum thawing layer is discussed in terms of chemical and biological safety in the area of human activities. The layout for glaciers protection has also been considered. Estimated cost of the concept implementation (ca. 180 $/m^2$) is just slightly higher than for widely used thermosyphons while providing much better technical performance and unique possibilities for revenue by selling (not wasting) heat and excessive electricity up to 70 $/(m^2 yr)$.

## 1. Introduction

* corresponding author: yagor@bmstu.ru, +7 903 786 9566

Permafrost is considered to be a part of the upper crustal layer characterized by a subzero temperature of rocks and / or soil for two or more years and the absence of seasonal thawing. Permafrost underlies 24% of land surface in Northern Hemisphere (about 22.8 million $km^2$), including more than 11 million $km^2$ in Russia, giving ca. 65% of the country's territory [1]. The general trend for global temperature increase leads to permafrost thawing.

Global warming is currently most pronounced in the Arctic, leading to up to 0.075°C/year air and 0.1°C/year soil temperature increase [2-4]. In the near future, soil surface temperature within the permafrost areas is going to increase by up to 2.3°C, leading to thawed layer depth increase by 33% and permafrost region edge shift by 50–600 km north [5]. Permafrost thawing is normally accelerated under constructions leading to deformations [6], and now it starts to cause disasters, e.g., recent 20 000 t diesel fuel spill in Norilsk, Russia [7]. This problem becomes very urgent in Russia, and cost-effective measures to prevent the destruction of structures located in permafrost areas need to be developed. Thawing also happens to glaciers, so extraordinary measures are implemented to stop this process at some sites [8], in part, to prevent disasters that could be easily induced at metastable balance interruption, e.g., by an earthquake.

Construction on permafrost soil has always been full of challenges [9]. Most were resolved at reasonable cost, to large extent, using thermosyphons for enhanced soil freezing during winter time so it is not considerably thawed during summer [10]. Now, the cost of sufficient solutions increases drastically, as well as need for constant monitoring of constructions and underlying soil state. This problem becomes particularly pronounced for linear infrastructural objects such as motorways [11], railroads [12], pipe [10] and power [13] lines connecting sparse settlements across the wilderness. To prevent disasters, chilling units accompanied by power plants, fuel tanks, and staff have to be brought in the most critical cases.

Basically, permafrost thawing happens as winters become milder and shorter while summers become hotter and longer. Particularly for this reason, widely used thermosyphons are currently operated out of designed conditions, so are not capable of keeping soil frozen. Industrial users of thermosyphons have measured actual performance of those to be just a half of that used in calculations. Moreover, there are operational data showing ca. 30% of thermosyphons can be damaged at mounting and up to 50% more breakdown within a decade.

Noteworthy, major research trends in this area are related to precise soil thermal state simulation rather than development of new or optimized methods of thermal stabilization [14, 15]. The need for innovation is characterized by the following situation. Thermosyphons had been proposed initially for soil freezing under the tips of foundation pillars, which are well bellow (ca. 6 m) the seasonally thawed layer (ca. 1 m). And since that worked well, nowadays, those are used widely, but not optimally, e.g., for thermal stabilization of the road embankments where soil of the upper layers is moving laterally mainly, so the effect of deep layers freezing is rather poor. Some reconsideration has started recently, and a combination of sloped, or flat looped thersmosyphons [16] and heat insulation of the surface is used to prevent upper soil layers thawing [17]. However, this combination of two methods, the complexity of assembly, and delivery costs for high volumes of heat insulation nearly double the cost of the solution compared to traditional vertical thermosyphons.

Heat is supplied to soil via convection at natural air motion, solar radiation absorption, and also comes with liquid precipitation. There are solutions to minimize this heat flow: solar reflectors and shields [18], heat insulation [12] and waterproof layers [11]. Even such an exotic way as grazing livestock is considered to suppress snow accumulation, and so getting better soil freezing in winter [19]. But obviously neither of these passive methods is capable of complete ambient heat rejection. Active cooling methods are well developed in general, but demand power supply, so are very expensive. However, sufficient energy could be obtained at right-of-way land of linear infrastructure objects using solar power plants [20]. Recently, suggestions to use solar powered devices for active thermal stabilization of permafrost railroads embankments have appeared [21, 22]. Work [22] is the closest to the suggested concept, but still relies on in depth cooling, does not suggest significant shielding of solar radiation and using benefits of grids operation (such as electric and cooling energy redistribution and heat utilization rather than wasting it).

Preliminary results showing potential capability of keeping soil frozen using local resources have been published earlier [20, 21]. The concept has been patented (patent RU 2 748 086 C1, ePCT application RU2021/050325) and proven experimentally [23]. In this article, we are presenting more detailed calculations results at different climatic conditions, compare those to traditional methods, and discuss possible technical layouts for a sustainable thermal stabilization system.

## 2. Theory

The main idea of our concept is to make solar radiation cool rather than heat the soil, for this:

1) soil has to be shielded from solar radiation and precipitation to the maximum extent, since the snow does not reach the ground either, it is better frozen in winter, and convection in summer is also partially suppressed;
2) better than just shielded, solar energy should be converted to electricity using photovoltaics (PV), thermovoltaics or their combination as well as collected in thermal form using vacuum pipes or plate collectors to drive the chiller;
3) regardless of which cooling principle is used (vapor compression [22], magnetocaloric [24], sorption [25]), the devices use either electricity or both electricity and heat harvested at previous step;
4) cooled near-surface layer should be organized to make a barrier for ambient heat penetration in depth rather than cooling huge volumes of soil as with thermosyphons, for that, probes are placed parallel to the ground surface within the natural thawing layer.

The general layout for this concept implementation is presented in Figure 1. All needed technologies are well developed and just need to be combined in the most efficient way. Thermal inertia of the soil makes the inconvenience associated with the intermittent power supply from renewable energy sources negligible. The system does not need to be operated in winter since natural cold is doing its work.

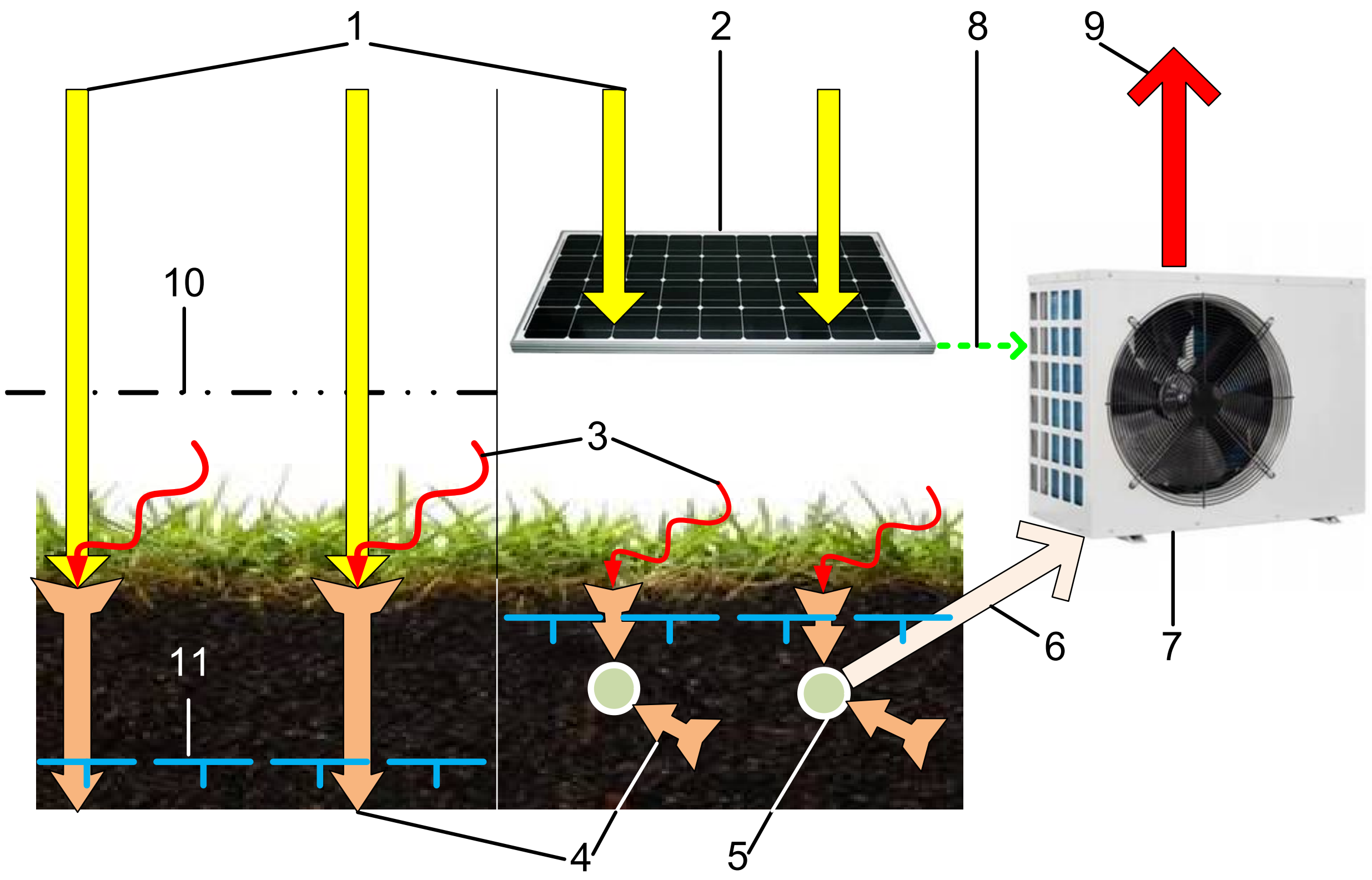

Figure 1. General schematic of a concept (right) compared to natural conditions (left): 1 – solar radiation; 2 – solar energy converter; 3 – convective heat flux; 4 – conductive heat flux; 5 – ground probes; 6 – heat sink from ground probes; 7 – chiller; 8 – converted solar energy; 9 – resulting heat sink; 10 – snow cover height; 11 – thawing layer boundary.

## 3. Calculations

Using QGIS v.2.18 open software, we have developed a geographic information system (GIS) containing layers (Figure 2) with data for: solar irradiation (global (GHI) and direct (DHI) on horizontal surface, monthly, annual) [26], air temperature (monthly) [26-28], wind speed (monthly) [26], soil temperature (different depths, monthly) [28], permafrost extension (by types) [29], railroads (by electrification type). Using these data, we have picked three test sites with rather different conditions: Norilsk (N 69.33°, E 88.21°), Yakutsk (N 62.03°, E 129.74°), and Chita (N 52.03°, E 113.5°). The former two are close to the terminal stations, and new railroads are going to be built there soon across continuous and slightly discontinuous permafrost. The latter is at the intensely used Trans-Siberian railroad, where traffic limitations occur due to the discontinuous and sporadic permafrost thawing. Since Melnikov Permafrost Institute of the Russian Academy of Sciences is located in Yakutsk, the most detailed data exist for this site; so it was used for most tests and preliminary evaluations.

For thermal calculations, we have used Frost-3D Multi-Core GPU software (https://frost3d.ru/eng/) which has been specially developed for permafrost soil calculations, its results validity has been verified by analytical solutions [30] and practical applications, compared to the most popular FEM packages [31], and conformity to relevant national and corporate (Gazprom) construction regulations certified. We realize more accurate (and inevitably more complicated) calculation methods might exist, but our aim at this stage was not to evaluate temperature distributions precisely, but to see if our concept is viable and to what extent.

Newton's law of cooling is used in the package being used; with third-type boundary conditions [32]. Albedo values taken: 0.75 for soil, 0.5 for concrete pavement, 0.1 for snow. Under solar shields, solar radiation was taken as 0.05 of GHI, wind speed as 0.3 of the reference value (measured at 10 m above soil surface). We calculated the convective heat transfer as $h = 6.16+4.19u$, where $u$ is the effective wind speed (that is already built into the software). Snow layer thermal resistance was also considered as wells as snow properties alteration over time (according to Russian building regulations SP 25.13330.2012).

The calculation area and soil properties are presented in Figure 3 and Table 1; climatic data used for initial conditions description in the software are presented in Tables 2–4. Time increment of 1 month was taken. September 2015 was taken as a start point for calculations opposed to January, to have soil unfrozen. Temperature distribution graphs are provided for the embankment axis. Initially, we have calculated temperature distributions for 5 years at natural conditions to see whether heat fluxes are balanced (Figure 4a). In case misbalance was observed, we adjusted soil layers composition since there are no precise data for this parameter for the exact soil temperature reference site. We assumed the results to be satisfactory if calculated temperature difference did not exceed 1°C/yr for any month and depth (Figure 4a). Then we have considered cases with sun shields, heat insulation, thermosyphons, sun shields + ground probes, sun shield + thermosyphons presence all applied to the North–South railroad embankment. Calculations were performed for 5 subsequent years to see when the new equilibrium state is reached.

We have considered the case of vapor compression heat pump powered by PV as the easiest to implement on site. Electric power capacity was evaluated using GHI and PV conversion efficiency of 10% (intentionally taken lower than most commercial products state for standard test conditions (STC)). Heat pump cooling capacity was evaluated by multiplication of the electric power capacity by cooling energy efficiency rate (EER, can be estimated by coefficient of performance COP usually specified for the heat pumps as EER=COP–1), dependent on ambient air temperature $t_a$ (°C) as EER = 4.8 – 0.12$t_a$ (this performance is also not the best available on market). For calculations, it affected heat exchange rate α inside the ground probes to match the available cooling capacity. Mid summer drop in EER was compensated to some extent by higher solar irradiation, and average heat transfer in ground loop was 45.7 W/(m$^2$K). These figures are reached for PV power achieved for the equal area covered with ground probes and PV panels. In our calculation area (Figure 3), solar panels are placed outside the embankment too, providing extra shielding to the soil and energy for other needs. Altso this energy could be used to power heat pumps leading to cooling capacity increase, e.g., when the upper part of the embankment is not shielded for traffic needs. Temperature of the liquid inside the ground probes changed from –10°C at the inlet to –4°C at the outlet. Once calculated soil temperature went below it, the heat pump was considered as switched off, i.e., α was zeroed. Ground probes were placed at 20 cm depth with 20 cm between the axes of 25 mm outer diameter, 2 mm thick wall low density polyethylene pipes.

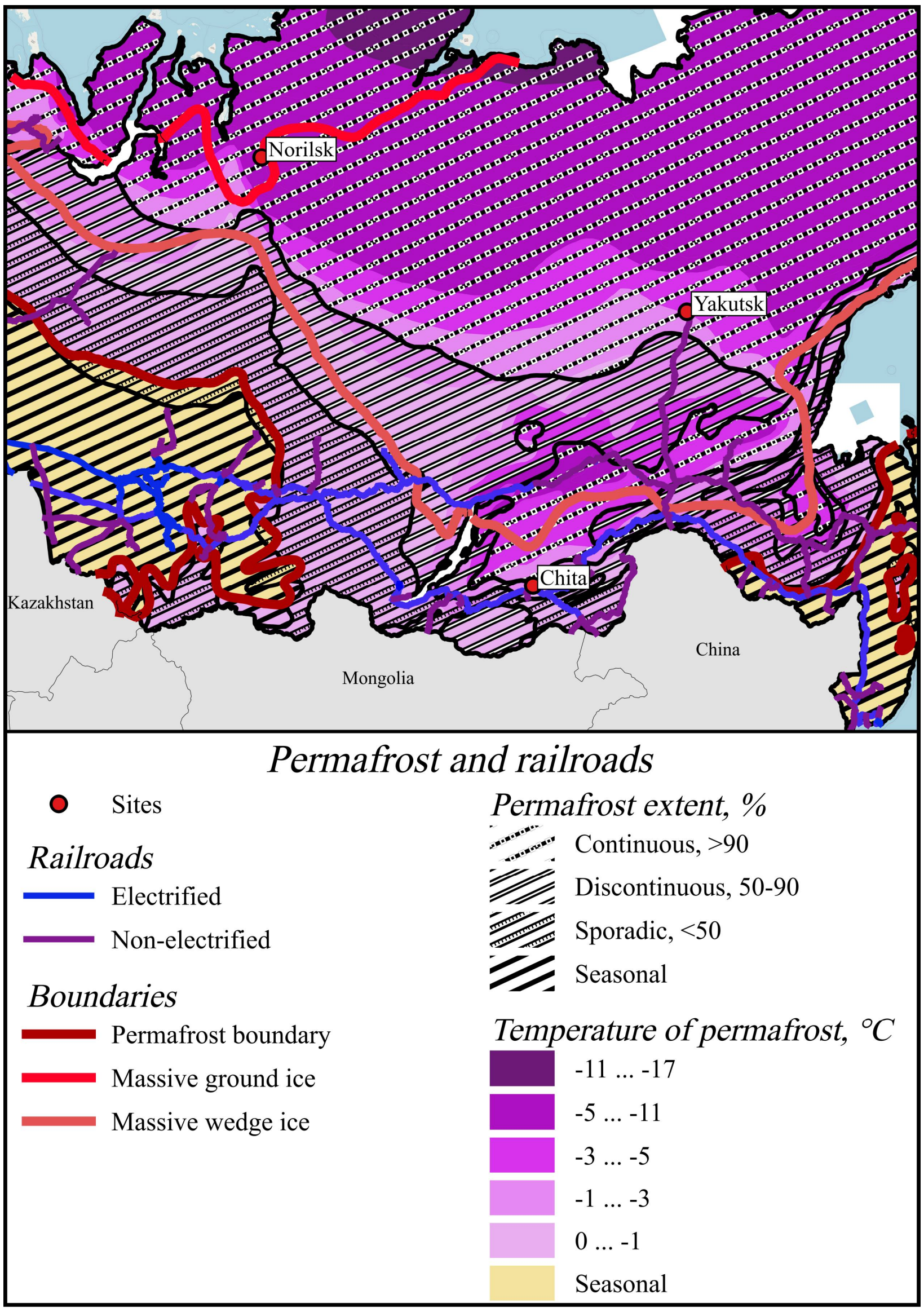

Figure 2. Considered sites and their position regarding to permafrost and railroads.

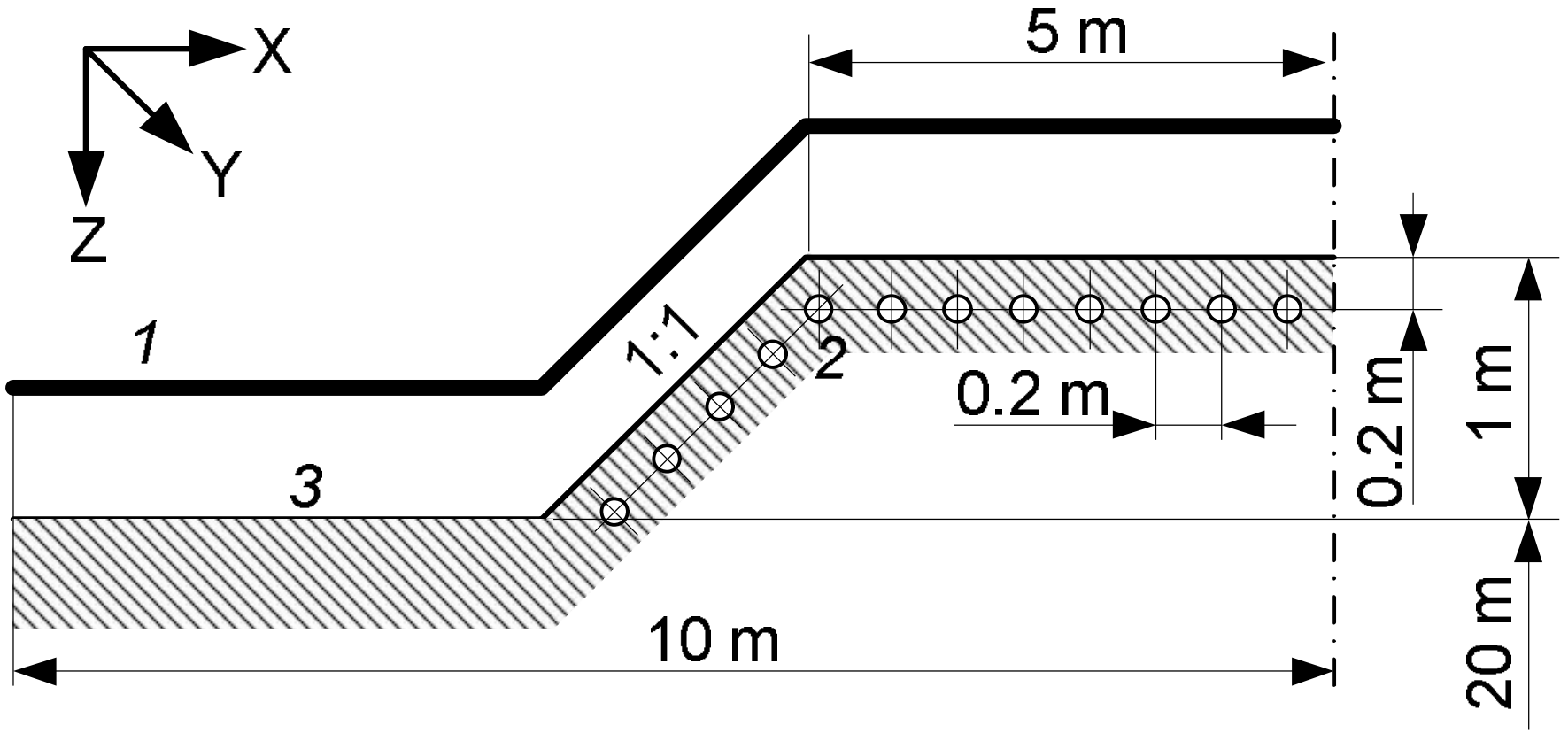

Figure 3. Calculation area and mesh characteristics: 1 – solar panels, 2 – ground probes, 3 – soil surface (number of cells – 48951; number of cells on the X axis – 441; number of cells on the Y axis – 1; number of cells on the Z axis – 111).

Table 1. Soil layers composition.

| Norilsk | | | | Yakutsk | | | | Chita | | | |
|---|---|---|---|---|---|---|---|---|---|---|---|
| Soil | Layer thickness, m | Moisture | Phase transition temperature, °C | Soil | Layer thickness, m | Moisture | Phase transition temperature, °C | Soil | Layer thickness, m | Moisture | Phase transition temperature, °C |
| Peat | 0.5 | 3.47 | 0 | Peat | 0.3 | 3.47 | –0.13 | Sandy loam | 2.3 | 0.1 | –0.2 |
| Clay loam | 1 | 0.25 | –0.21 | Sand | 1 | 0.4 | –0.11 | Sand | 1.2 | 0.2 | –0.12 |
| Sandy loam | 19 | 0.5 | –0.16 | Sandy loam | 5 | 0.56 | –0.16 | Gravel | 16.5 | 0.2 | 0 |
| | | | | Clay loam | 14 | 0.66 | –0.21 | | | | |

Table 2. Climatic data for Norilsk.

| Month | Solar irradiation. kWh/(m$^2$*mon.), (averaged W/m$^2$) | Mean air temp., °C | Mean wind speed, m/s | Snow layer, cm | Soil temperature at depths, °C | | | | |
|---|---|---|---|---|---|---|---|---|---|
| | | | | | 20 cm | 40 cm | 80 cm | 160 cm | 320 cm |
| 1 | 0.31 (0.417) | –26.2 | 5.1 | 21.33 | –4.1 | –2.6 | –1.7 | –0.3 | –1.1 |
| 2 | 8.68 (12.9) | –24.2 | 4.8 | 21.00 | –7.6 | –5.4 | –4.0 | –0.6 | –1.0 |
| 3 | 42.16 (56.67) | –19.2 | 4.3 | 21.67 | –6.2 | –4.8 | –4.2 | –1.4 | –0.9 |
| 4 | 97.2 (135) | –14.4 | 4 | 21.33 | –5.3 | –4.6 | –4.2 | –2.1 | –1.0 |
| 5 | 146.9 (197.5) | –5.3 | 3.7 | 12.67 | –0.3 | –2.2 | –2.3 | –2 | –1.3 |
| 6 | 155.1 (215.4) | 4.1 | 3.6 | 0.00 | 0.1 | –0.6 | –0.9 | –1.5 | –1.5 |
| 7 | 152.8 (205.4) | 12.7 | 3.8 | 0.00 | 9.7 | 5.3 | –0.1 | –1.0 | –1.4 |
| 8 | 102.6 (137.9) | 10 | 3.8 | 0.00 | 6.4 | 4.8 | 3.3 | –0.4 | –1.3 |
| 9 | 53.4 (74.17) | 1.7 | 4 | 0.00 | 4.8 | 4.5 | 3.7 | 0 | –1.3 |
| 10 | 19.53 (26.25) | –10.6 | 4.4 | 6.33 | 1.7 | 1.9 | 1.8 | 0.2 | –1.0 |
| 11 | 1.8 (2.5) | –20.9 | 4.8 | 13.33 | –0.9 | –0.2 | 0 | 0 | –1.0 |
| 12 | 0 (0) | –24.9 | 5 | 18.00 | –7.8 | –4.7 | –1.9 | –0.3 | –1.0 |

Table 3. Climatic data for Yakutsk.

| Month | Solar irradiation. kWh/(m$^2$*mon.), (averaged W/m$^2$) | Mean air temp., °C | Mean wind speed, m/s | Snow layer, cm | Soil temperature at depths, °C | | | | |
|---|---|---|---|---|---|---|---|---|---|
| | | | | | 20 cm | 40 cm | 80 cm | 160 cm | 320 cm |
| 1 | 8.06 (10.83) | –36.5 | 1.5 | 26.1 | –7.4 | –6.4 | –3.7 | –0.2 | –0.6 |
| 2 | 22.68 (33.75) | –35.1 | 1.5 | 31.4 | –8.5 | –7.4 | –5.3 | –0.9 | –0.5 |
| 3 | 86.18 (115.8) | –15.4 | 1.9 | 31.4 | –6.9 | –6.3 | –4.9 | –1.7 | –0.5 |
| 4 | 134.1 (186.3) | –2.1 | 2.7 | 15.3 | –4.3 | –3.9 | –3.3 | –1.7 | –0.5 |
| 5 | 163.4 (219.6) | 9.5 | 3.5 | 0 | 4.0 | 1.7 | –1.5 | –1.3 | –0.5 |
| 6 | 175.8 (244.2) | 16.1 | 3.1 | 0 | 12.8 | 10.5 | 4.6 | –0.9 | –0.5 |
| 7 | 166.5 (223.8) | 22.5 | 2.9 | 0 | 20.8 | 17.3 | 10.7 | 0.6 | –0.5 |
| 8 | 132.1 (177.5) | 17.4 | 2.8 | 0 | 15.4 | 14.6 | 12.3 | 5.2 | –0.5 |
| 9 | 72.3 (100.4) | 3.8 | 2.6 | 0 | 6.0 | 6.3 | 6.4 | 4.8 | –0.5 |
| 10 | 37.82 (50.83) | –5.1 | 2.6 | 5.6 | 0.3 | 0.9 | 1.9 | 2.2 | –0.3 |
| 11 | 12.6 (17.5) | –26.5 | 1.9 | 17.1 | –2.5 | –1.6 | –0.6 | 0.3 | –0.5 |
| 12 | 4.34 (5.83) | –36.1 | 1.5 | 23.4 | –4.8 | –3.4 | –1.2 | –0.2 | –0.5 |

Table 4. Climatic data for Chita.

| Month | Solar irradiation. kWh/(m$^2$*mon.), (averaged W/m$^2$) | Mean air temp., °C | Mean wind speed, m/s | Snow layer, cm | Soil temperature at depths, °C | | | | |
|---|---|---|---|---|---|---|---|---|---|
| | | | | | 20 cm | 40 cm | 80 cm | 160 cm | 320 cm |
| 1 | 33.48 (45) | –27.2 | 1.4 | 18.1 | –16.4 | –11.9 | –7.7 | –2.4 | 1.0 |
| 2 | 59.92 (89.17) | –17.9 | 1.8 | 13.1 | –16.6 | –13.2 | –9.6 | –4.9 | 0 |
| 3 | 112.8 (151.7) | –6.6 | 3 | 3.3 | –6.2 | –5.5 | –5.3 | –4.6 | –0.4 |
| 4 | 145.8 (202.5) | 4.8 | 4.1 | 0 | 5.2 | 2.7 | –0.8 | –1.8 | –0.7 |
| 5 | 180.1 (242.1) | 9.8 | 4 | 0 | 12.0 | 8.6 | 5.0 | 0.6 | –0.2 |
| 6 | 184.2 (255.8) | 18.6 | 3.1 | 0 | 20.9 | 15.9 | 10.6 | 4.5 | 0 |
| 7 | 165.2 (222.1) | 19.1 | 2.6 | 0 | 19.6 | 17.7 | 14.0 | 8.4 | 3.1 |
| 8 | 138.6 (186.3) | 18.3 | 2.3 | 0 | 17.7 | 17.1 | 15.4 | 11.2 | 6.0 |
| 9 | 105.9 (147.1) | 7.3 | 2.8 | 0.6 | 9.1 | 11.2 | 11.8 | 10.8 | 7.4 |
| 10 | 73.16 (98.33) | 1.8 | 3 | 0 | 6.6 | 6.9 | 7.3 | 7.9 | 7.0 |
| 11 | 39.3 (54.58) | –14.7 | 2.7 | 0.5 | –10.1 | –3.8 | 0.6 | 3.9 | 5.4 |
| 12 | 25.73 (34.58) | –23.4 | 1.8 | 4.0 | –16.7 | –11.5 | –6.5 | –0.3 | 3.0 |

## 4. Results

### *4.1. Verification of calculations and traditional methods*

Figure 4a shows the year-to-year stability of the soil temperature at natural conditions to evaluate the stability of the calculation model. Figure 4b shows monthly distributions in 5 years after calculation start. Those were in good agreement with the existing reference data (Table 3). The next step was to evaluate solar panels shielding effect on the soil temperature distributions (Figure 5). It can be seen that surface temperature becomes significantly lower due to no snow cover under the shields. However, in-depth soil temperature is almost not affected. For this reason, maximum surface temperature is also decreasing, but to lower extent, and the thawing layer is reduced by nearly 0.5 m. Solar blinds have been used at the Baikal-Amur railroad to the south of Yakutsk; thawing layer was reduced from 4.0 m to 1.51 m in 5 years, correspondingly [33]. The effect of solar blinds is proportional to the radiation to convection heat flux ratio. For this reason, strong effect has been shown for Tibet [18]. But in some cases, 1.5 m thawed layer is still not appropriate.

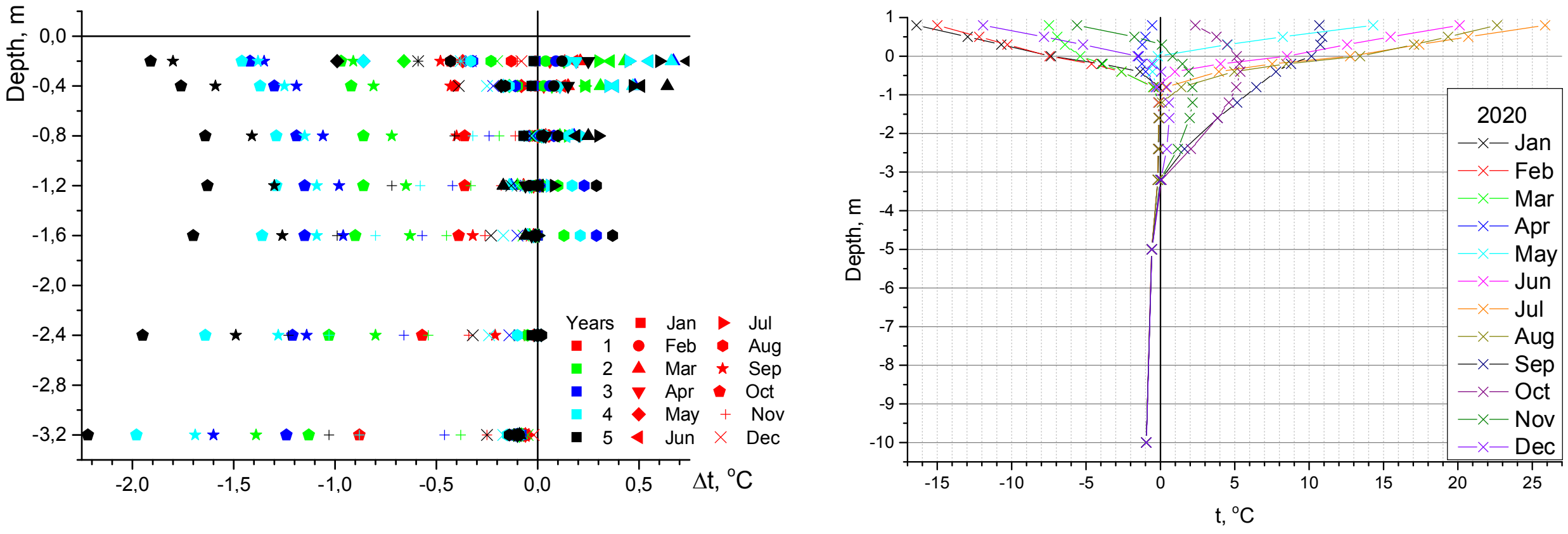

a b

Figure 4. Calculated soil temperature distribution at natural conditions to check for basic equilibrium in Yakutsk (a – difference to reference data over years after start of calculation; b – monthly distribution across the calculated embankment axis).

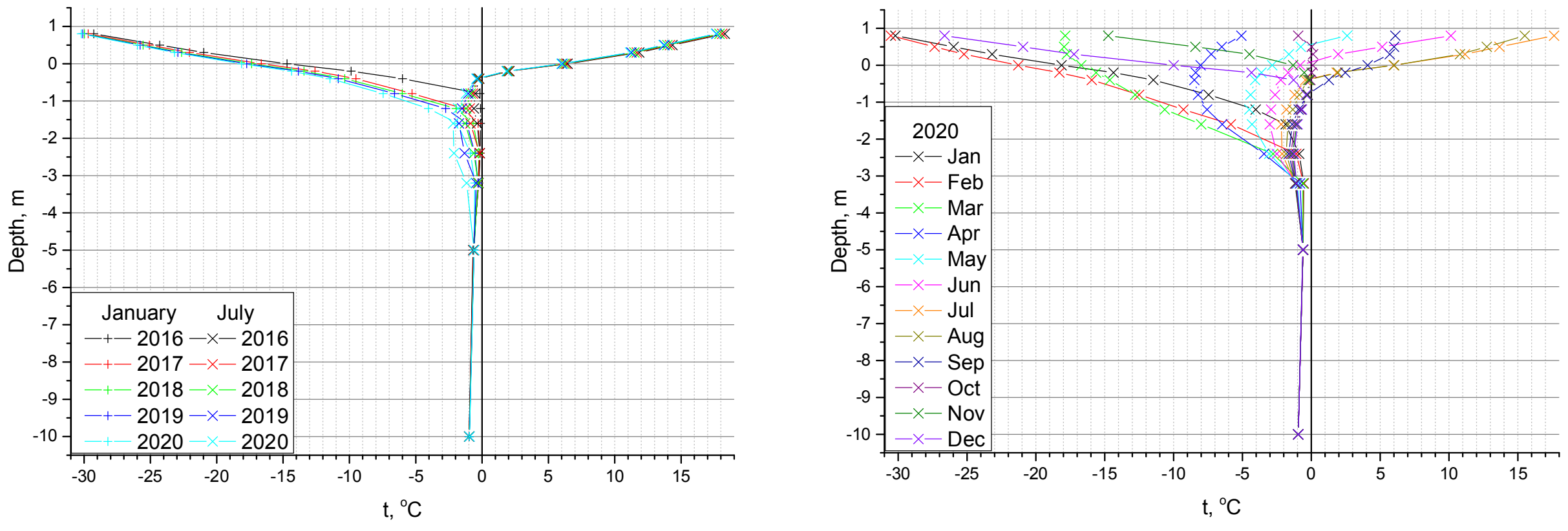

a b

Figure 5. Calculated temperature distribution under PV solar shields in Yakutsk (a – July and January in 2016–2020; b – monthly distribution in 2020).

Another way for shielding soil from the external heat is covering with the heat insulation layer. Crushed rock is a more usual and universal one, but polystirol sheets are also used for this. We have considered such way for Chita (Figure 6) since summer heating is rather high there, and convective part should also be rejected to get small thawing layer. However, such heat insulation prevents intense freezing in winter. For deep snow regions, it gives low additional effect, but snow cover is thin in Chita (Table 4). As a result, cooling in winter is very poor and highest temperatures are reached in September – October due to thermal conduction from the unshielded areas. Even though a thawed layer of several meters is still present, it exists for shorter time and is still thinner than at natural conditions. Another feature of this case is positive long-term temperature drift caused by low cooling in winter (Figure 6a), while all other cases are showing neutral or negative trend. This case is also sensitive to the season when the heat insulation is applied. If done in September, the soil is preserved in warm state and thawed deeper. If done in April, just after snow is gone, soil is preserved in the cooled state presented in Figure 6.

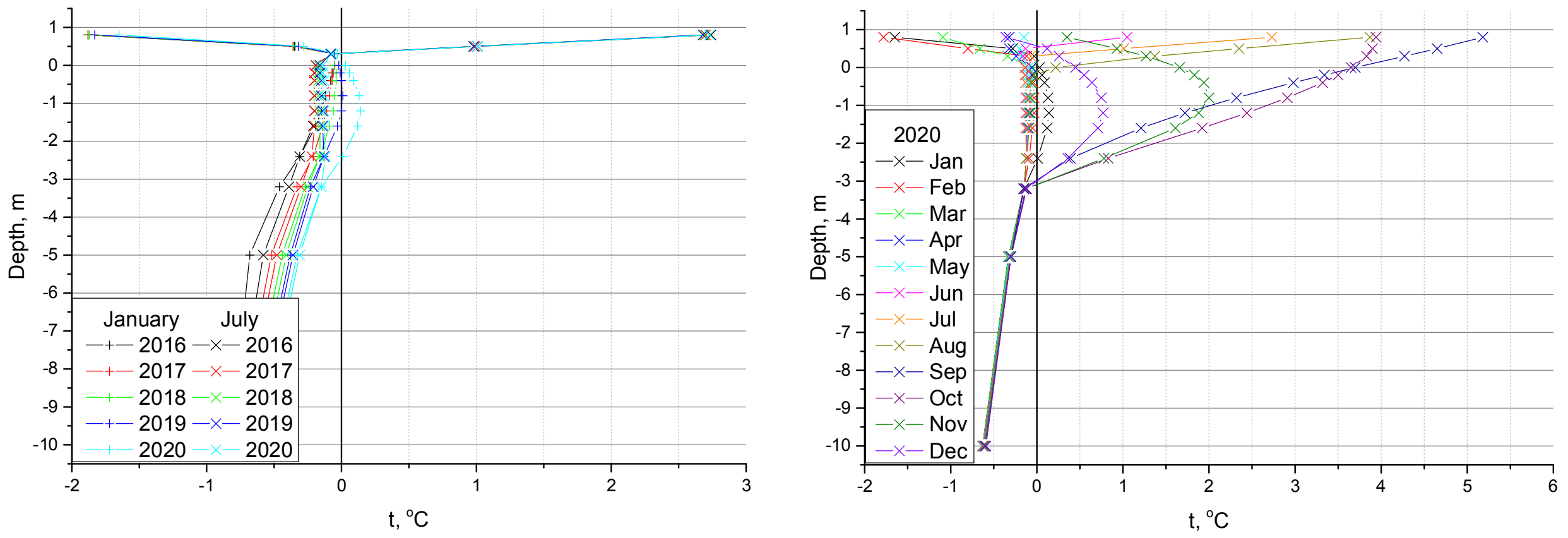

a b

Figure 6. Calculated temperature distribution under 50 mm polystirol heat insulation in Chita (a, b – see Figure 5).

Thermosyphons widely used for permafrost thermal stabilization give a similar effect for the embankment axis (Figure 7) compared to the solar blinds (Figure 5). While the latter are covering the embankment slopes uniformly, the former are forming cold spots, and their effect is decreasing with distance. If the evaporator part is vertical, the embankment axis can be too far to see the cooling effect, while frozen walls preventing soil movement are still formed aside, and deep layers are gradually cooling down. To make the embankment core frozen, sloped evaporators are used, but this complicates the thermosyphon design, logistics, and installation. The soil is still thawing down to 1.5 – 2.0 m since it is heated in summer, but this layer stays for a shorter time than at natural conditions. The time shift between artificial cooling and natural heating leads to the present issues in the warming climate. E.g., in hot summer after a warm winter, thawed gaps in the thermosyphon rows appear leading to wet soil movement through those bottlenecks. Noteworthy, dry and rocky soils characteristic for the southern edge of the permafrost extent are even less capable of such cold accumulation.

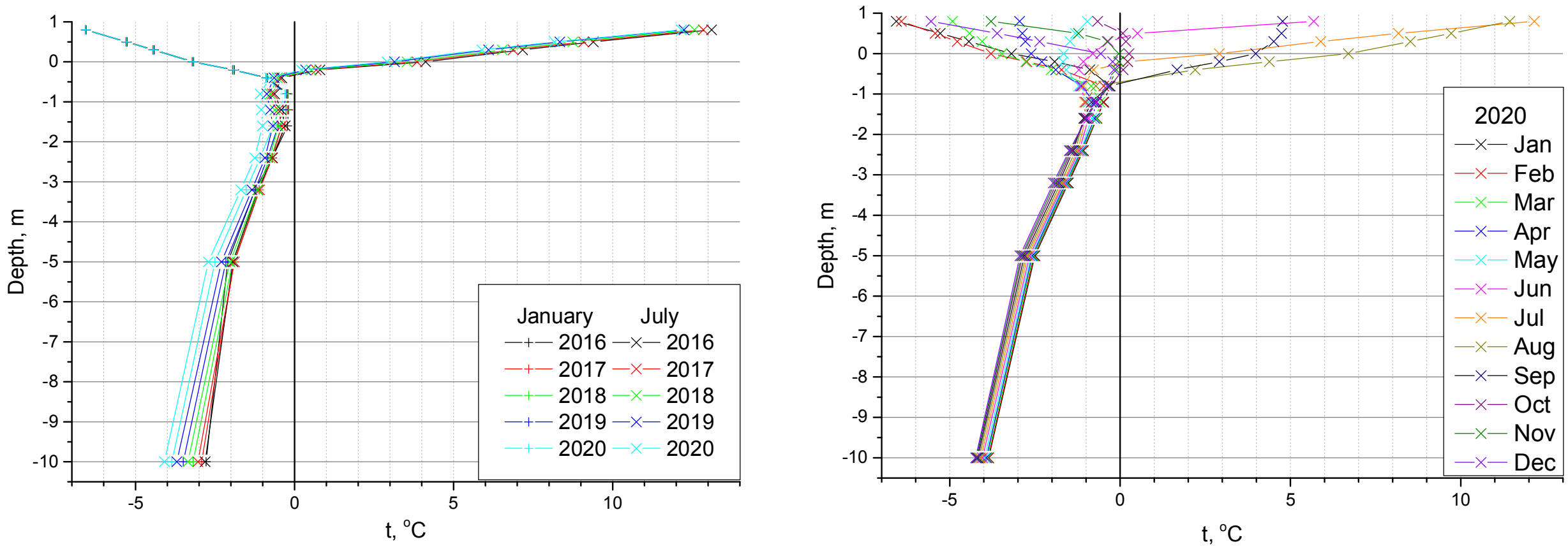

a b

Figure 7. Calculated temperature distribution under thermosyphons in Norilsk (a, b – see Figure 5).

To prevent this, the active cooling of thermosyphons is used in summer, solar powered chillers can be used to make such systems autonomous [22]. Or solar blinds could be combined with thermosyphons, but our calculations show (Figure 8) that despite certain effect in Norilsk, this way is still incapable of thawing layer significant reduction farther south. Neither does the combination of heat insulation with solar panels since the latter do not add anything sufficient to the shielding effect.

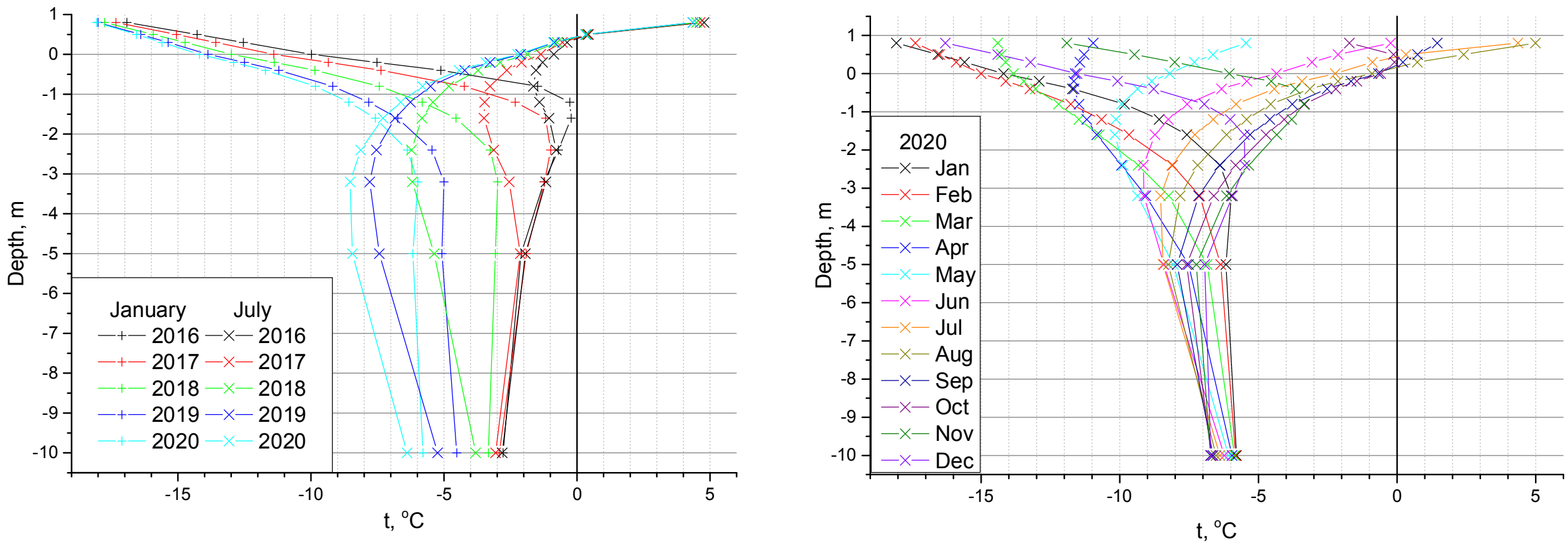

a b

Figure 8. Calculated temperature distribution under solar blinds combined with thermosyphons in Norilsk (a, b – see Figure 5).

*4.2. Implementation of the suggested approach*

Our approach suggests that cooling should be applied in summer not to the deep layers but to the soil surface – to prevent heat penetration completely. In terms of hardware, the system should be some similar to the one described in [22], but ground probes should be placed horizontally, like in the floor heating, and solar panels should cover most of the protected area possible. Figure 9 shows that almost no thawing layer is possible in Yakutsk conditions, and there is a remarkable negative temperature drift both in summer and winter even after 5 years of system operation. Figure 10 indicates clearly that heat does not penetrate beyond the ground probes even in the warmest conditions of considered, placing those closer to the surface can even eliminate thawing layer completely as soon as heat source is balanced with heat sink. Figure 10c is particularly interesting because the embankment stays frozen in summer while thawed layer is formed in-depth aside by long-distance thermal conduction.

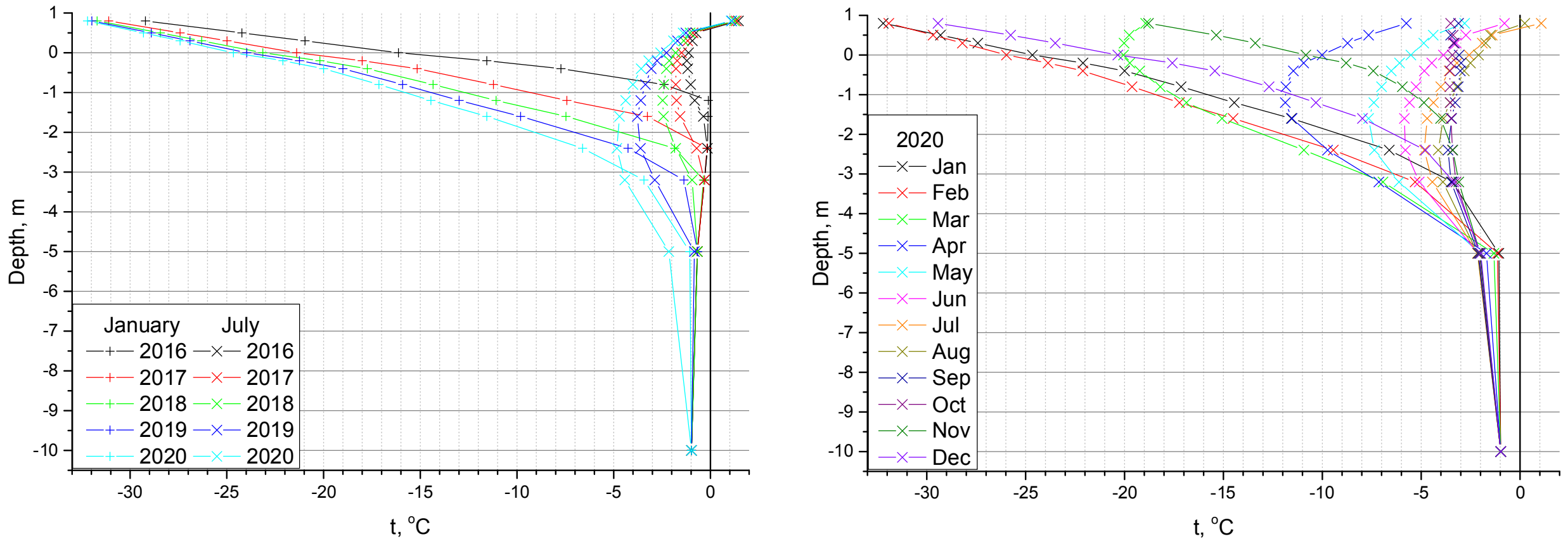

a b

Figure 9. Calculated temperature distribution under PV with near-surface ground probes in Yakutsk (a, b – see Figure 5).

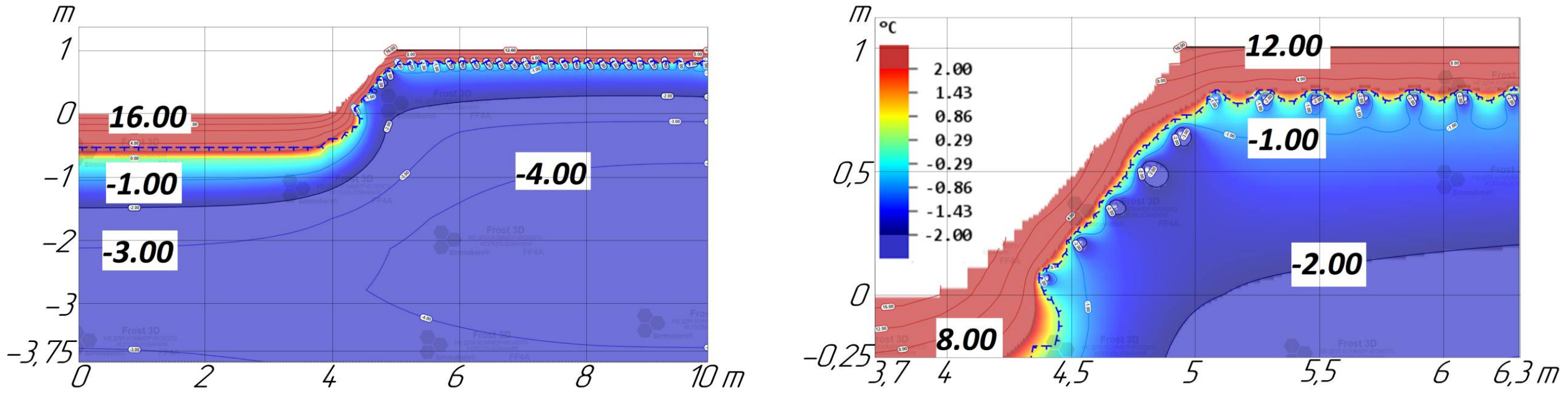

a

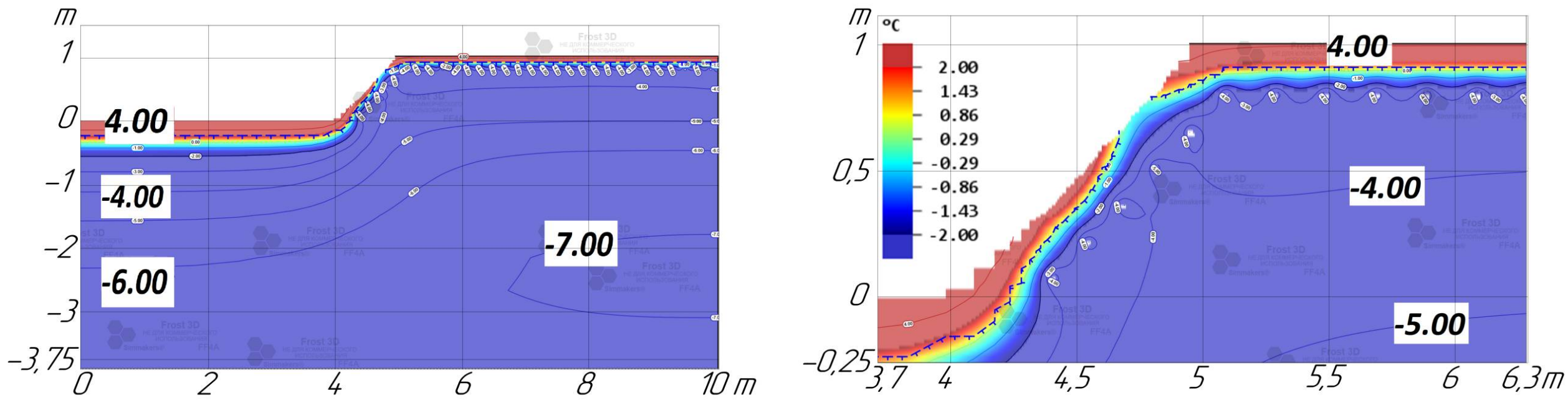

b

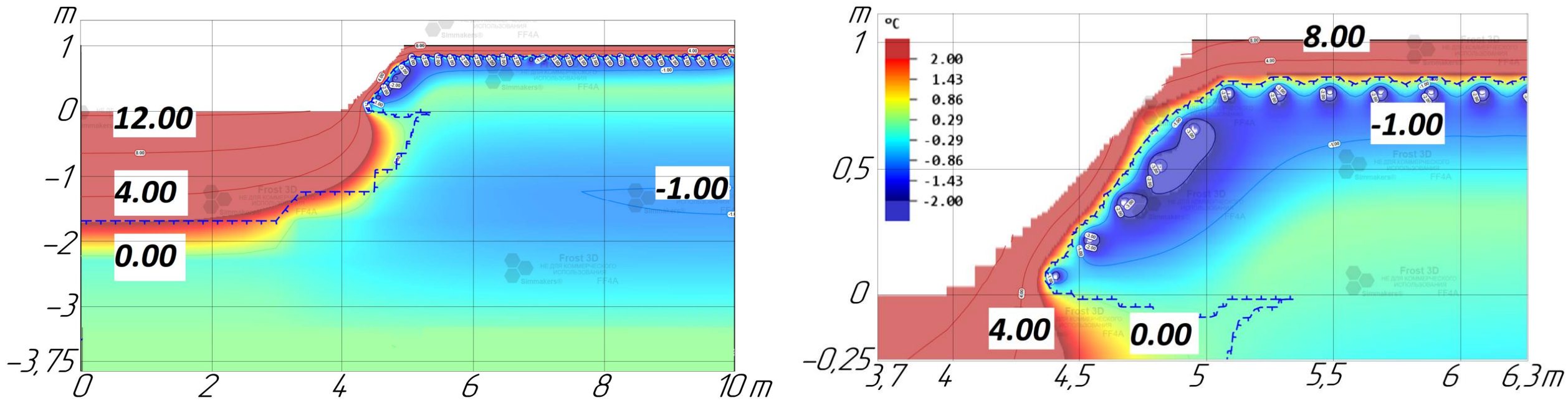

c

Figure 10. Calculated temperature maps under PV with near-surface ground probes in July 2020 (a – Yakutsk, b – Norilsk, c – Chita; left – general view, right – magnified upper edge of the embankment – the thermal shielding layer can be seen clearly).

There is a more exotic way to form a thermal shielding layer – spraying of artificial snow under solar panels. The application range for this way is more limited though. Of considered sites, it worked in Norilsk only leading to thawing layer of less than 0.5 m (Figure 11). This way is much more favorable for the alpine areas, where rocks and surface ice restrict using ground probes and convective heat flux is low. However, producing artificial snow suggests intense supply of finely filtered water. We have evaluated 6 cm thick artificial snow layer can be produced daily during months with positive mean air temperatures and based on that performed our calculations.

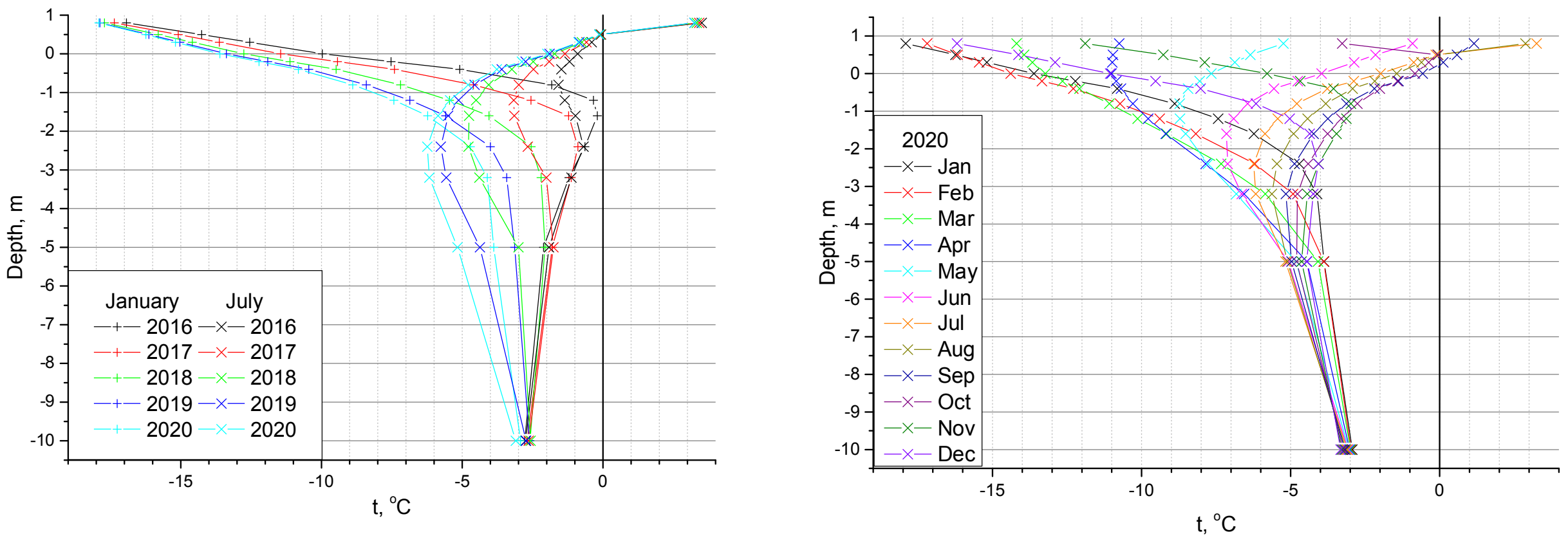

a b

Figure 11. Calculated temperature distribution under PV with artificial snow layer in Norilsk (a, b – see Figure 5).

### 5. Discussion

*5.1. Application to real constructions*

Even though sun elevation angles in the Arctic are low, we still suggest mounting solar converters parallel to the ground surface to provide the best shielding and to reduce wind loads. The exception relates to the buildings and fuel reservoirs, where vertical surface is even more desirable for the solar panels (if safety regulations are matched). In case of insufficient power supply, solar energy converters can cover the area bigger than what actually needs thermal stabilization, except extra power, those provide a buffer layer of shielded soil that is still cooler than at natural conditions. The example of such approach is presented in Figure 12. Without shielding the top of the embankment, heat flux is higher than sinking to the ground probes, this leads to deeper thawing. But side ground probes are forming frozen walls preventing thawed soil tangential movement. Dull weather does not suggest strong heating, but additional use of wind energy can be considered as a complementary source [34].

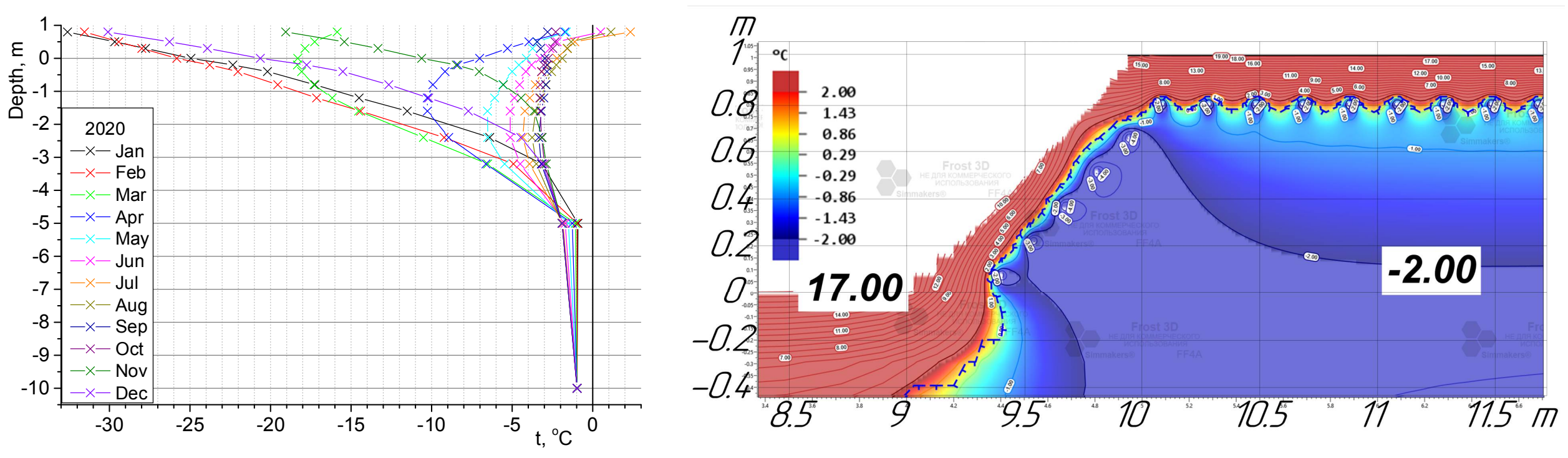

a b

Figure 12. Calculated temperature distributions at realistic road embankment (paved with concrete, no PV atop of it) in Yakutsk with 1.8x cooling capacity due to use of power from extra solar panels outside the embankment.

Ideally, to reduce digging, ground probes are to be placed as close to the surface as possible. But, in this case, heat flow is maximized and, to provide a uniform temperature field, probes have to be closer to each other. To some extent, this could be resolved by placing a heat insulation layer on top. But we suggest probes should be buried at 20 to 50 cm with about the same distance in between. Deeper deployed probes experience lower heat loads, and thermal inertia is higher, which is great for the cooling system (lower power demand, and so, lower cost). But that also means deeper thawing which might be bad for the protected object. This tradeoff should always be resolved individually to optimize the project cost and risks. This concept has been proven experimentally to be capable of preserving 26 cm deep near-surface shielding layer frozen even outside the permafrost area – at higher thermal loads [23]. Shallow horizontal ground probes are particularly useful in the regions with rocky general substrate, e.g., in the mountain areas. Thermal shielding layer could be particularly useful for saline soils since those are thawing deeper and need to be cooled to lower temperatures [35].

It has been shown that at permafrost thawing highest emission rates for $CO_2$ occur from 0–40 cm depths, for $CH_4$ – from 40–80 cm [36]. High abundances of carbon-cycling bacteria, fungi, and archaea corresponded to 0–40 cm depths. Release of hazardous chemical, e.g., mercury [37], has been reported too. So significant reduction of thawing layer and its temperature leads not only to better structure stability, but also prevents chemical and biological hazards in the area of human activities [38]. In critical cases, river and lake banks as well as coastal slopes can be also shielded using our approach to stop destruction, to prevent methane bubbling [39], thermokarst development and gas emission craters [40].

*5.2. Costs evaluation*

We have evaluated the basic cost of PV + vapor compression heat pump solution. PV panels current spot price is 0.171 \$/W [41], and reasonable conversion efficiency for this price is ca. 17% at STC, which gives 170 W/m$^2$ or 29 \$/m$^2$. PV panels represent ca. 40% of the solar power plant cost, so 75 \$/m$^2$ can be assumed. This cost could be reduced if DC driven compressors are used, so no invertors needed. Indeed, maximum power output for a horizontal panel at N 60° latitude (maximum sun elevation angle 53.44°) will be reduced to ca. 100 W/m$^2$, so this value is to be used for available cooling capacity calculation. High COP vapor compression pumps for a suitable temperature range cost ca. 1.0 \$/W$_e$, or 100 \$/m$^2$ (can be reduced, see below). Usually, horizontal ground probes installation cost depends on excavation works and is sufficient, but using cable laying machines in this case can reduce it to 2 \$/m$^2$ including pipe. So the whole suggested thermal stabilization system cost starts at 177 \$/m$^2$.

Thermosyphon cost is ca. \$330 and it freezes area of ca. 3 m$^2$, its mounting costs at least \$200, giving a total of at least 176 \$/m$^2$. For solar powered ground chillers [22], cost evaluation by the authors is ca. 125 \$/m$^2$, but it is provided for system components only, not including assembly and installation, which all can double the final price. Of course, these cost evaluations are coarse, but those show our solution might be economically competitive while providing cooling exactly when it is needed and with positive feedback to solar radiation, resulting in nearly guaranteed frozen state of the near surface soil layers. This feedback also resolves the problem of faster thawing at southern slopes – at more intense solar flux.

The reduction of construction and maintenance costs at guaranteed permafrost thermal stabilization needs special consideration. Rough estimates show that these savings are at least half of the suggested solution cost, so making it eventually cheaper than traditional approaches. In case energy pumped out of the ground is supplied for heating rather than released to the atmosphere, additional income is generated. For example, ca. 0.3 Gcal/($m^2$yr) output in Norilsk, at local rates for heating reaching 250 \$/Gcal for isolated small communities [42], results in saving up to 75 \$/($m^2$yr). Even though half to quarter of this are more usual rates for comparatively developed sites in that region (and tenth for the city), the suggested system could be paid off just in this way during its lifetime.

*5.3. Possible optimization*

The electricity is the most universal form of energy to be used in devices, but photovoltaic (and thermovoltaic to even greater extent) conversion efficiency is rather low if cheaper products are considered. While electricity-driven refrigeration units provide the highest COP. As for any system, cooperation of elements at certain conditions should be considered. For example, we can consider PV combined with a vapor compression cooler or a vacuum tube collector – with an absorption cooler. Commercial PV has realistic efficiency of 15%, and vacuum tube solar collectors have 60%. A good vapor compression heat pump would give us COP of 4.0 while absorption pump (uses mainly heat and needs electricity for circulation pumps, fans and other service) – just 2.0.

Multiplication of performance for these two cases gives approximate overall use of solar energy of 60% and 120%. This still does not mean the latter configuration is better. While vacuum tube pipes convert direct solar radiation only, some PV (a-Si, CIGS) can deal well with the diffused radiation too. In reverse, at higher ambient temperature, vacuum tube efficiency grows while it goes down for PV (except some non-mass production materials). Vapor compression coolers specific cost is lower than absorption ones at cooling capacity below ca. 2 MW. COP of vapor compression heat pumps can be improved if those are working in invertor mode powered with DC current from solar panels without multiple conversion DC/AC/DC/AC as in conventional grid powered invertor units. Invertor mode allows the load to follow the solar energy output reducing need in expensive buffer energy storage and capacity reserve for peak currents drastically. All this is to show a number of tradeoffs to be resolved in each certain case.

At first glance, bulk energy storage (except minimum needed for control units operation and slight damping of power supply intermittence) will not be paid off. Since in this case, the soil is the major and free storage of the demanded thermal energy. What could be optimized is PV to cooling unit peak capacity ratio. Capacity factor for PV part depends a lot on local conditions, and for horizontal panels is basically limited by the site latitude (on average, will not exceed 18%; in this case, it should be calculated for active season rather than the whole year). As it was shown above, cooling unit specific price is significantly higher than PV, so its capacity factor is worth optimization. For that, cooling unit operation should be uniform; could be supported by driving energy storage. Bulk heat storage is considerably cheaper than electric, which gives another advantage for solar thermal + absorption cooling configuration. But in case of excessive electric energy generation, it could be sold out where applicable, so making investment in PV power more cost effective than in energy storage.

## 6. Conclusions and future work

A sustainable way to reduce thawing layer depth to the first decimeters or eliminate it completely is presented for the first time. The concept combines previously known ways of solar radiation and precipitation shielding, solar powered cooling, and a novel approach suggesting forming a near-surface thermal shielding layer to prevent heat penetration in depth when it is actually supplied to the soil surface. That is suggested to be formed by shallow horizontal ground probes or, in some cases, by artificial snow. The cooling capacity of this system has a positive feedback to the solar irradiation – the main component of soil heating – so ensuring extra cooling capacity at increased thermal loads.

In addition to infrastructure and buildings protection, this concept could be used to prevent greenhouse gases release, and chemical and biological pollution from thawing permafrost in the areas of human activities at least. We suggest this solution could be highly demanded for constructions life cycle contracts. The evaluated cost of its implementation (180 $\$/m^2$) is comparable to widely used thermosyphons while much better technical effect (by thawing layer reduction to first decimeters), future cost reduction (due to long-term negative trends in prices for PV modules and heat pumps), and unlike any other methods, possibilities for revenue by selling (not wasting) heat and excessive electricity up to 70 $\$/(m^2yr)$ are expected.

**Acknowledgements**

We are grateful to Simmakers Ltd. for providing free remote access to the Frost 3D package to ESS. Work of EYL has been supported by the Russian Ministry of Science and Higher Education [state assignment No. 0705-2020-0046]. Equipment of the “Beam-M” facility (Bauman Moscow State Technical University) has been used in this work. Authors are grateful to J.P. Clements for English proofreading.

## References

[1] O.A. Anisimov, Continental permafrost, Roshydromet, Moscow, 2012 (in Russian).
[2] L. Chen, W. Yu, Y. Lu, W. Liu, Numerical simulation on the performance of thermosyphon adopted to mitigate thaw settlement of embankment in sandy permafrost zone, Applied Thermal Engineering, 128 (2018) 1624-1633. doi: 10.1016/j.applthermaleng.2017.09.130.
[3] E. Post, R.B. Alley, T.R. Christensen, M. Macias-Fauria, B.C. Forbes, M.N. Gooseff, A. Iler, J.T. Kerby, K.L. Laidre, M.E. Mann, J. Olofsson, J.C. Stroeve, F. Ulmer, R.A. Virginia, M. Wang, The polar regions in a 2°C warmer world, Science Advances, 5 (2019) eaaw9883. doi: 10.1126/sciadv.aaw9883.
[4] J. Van Huissteden, Thawing permafrost: Permafrost carbon in a warming arctic, 2020. doi: 10.1007/978-3-030-31379-1.
[5] A.V. Pavlov, G.F. Gravis, Permafrost and modern climate, Russian academy of science: Priroda, 1016 (2000) 10-18 (in Russian).

[6] C. Yuan, Q. Yu, Y. You, L. Guo, Deformation mechanism of an expressway embankment in warm and high ice content permafrost regions, Applied Thermal Engineering, 121 (2017) 1032-1039. doi: 10.1016/j.applthermaleng.2017.04.128.
[7] BBCNews, Arctic Circle oil spill: Russian prosecutors order checks at permafrost sites, https://www.bbc.com/news/world-europe-52941845, accessed: 2020.09.17.
[8] C. Harvey, Can we refreeze the Arctic? Scientists are beginning to ask, https://www.eenews.net/stories/106007550, accessed: 2020.09.17.
[9] S.W. Muller, H.M. French, F.E. Nelson, Frozen in time: Permafrost and engineering problems, 2008. doi: 10.1061/9780784409893.
[10] G. Li, F. Wang, W. Ma, R. Fortier, Y. Mu, Z. Zhou, Y. Mao, Y. Cai, Field observations of cooling performance of thermosyphons on permafrost under the China-Russia Crude Oil Pipeline, Applied Thermal Engineering, 141 (2018) 688-696. doi: 10.1016/j.applthermaleng.2018.06.005.
[11] D. Yinfei, W. Shengyue, W. Shuangjie, C. Jianbing, Cooling permafrost embankment by enhancing oriented heat conduction in asphalt pavement, Applied Thermal Engineering, 103 (2016) 305-313. doi: 10.1016/j.applthermaleng.2016.04.115.
[12] J. Luo, F. Niu, M. Liu, Z. Lin, G. Yin, Field experimental study on long-term cooling and deformation characteristics of crushed-rock revetment embankment at the Qinghai–Tibet Railway, Applied Thermal Engineering, 139 (2018) 256-263. doi: 10.1016/j.applthermaleng.2018.04.138.
[13] T. Wang, G. Zhou, D. Chao, L. Yin, Influence of hydration heat on stochastic thermal regime of frozen soil foundation considering spatial variability of thermal parameters, Applied Thermal Engineering, 142 (2018) 1-9. doi: 10.1016/j.applthermaleng.2018.06.069.
[14] X. Kong, G. Doré, F. Calmels, Thermal modeling of heat balance through embankments in permafrost regions, Cold Regions Science and Technology, 158 (2019) 117-127. doi: https://doi.org/10.1016/j.coldregions.2018.11.013.
[15] X. Kong, Development of design tools for convection mitigation techniques to preserve permafrost under northern transportation infrastructure, in, Vol. Ph.D., Universite Laval, Québec, 2019, pp. 185.
[16] I. Holubec, Flat Loop Thermosyphon Foundations in Warm Permafrost in, I. Holubec Consulting Inc., 2008.
[17] Y. Mu, G. Wang, Q. Yu, G. Li, W. Ma, S. Zhao, Thermal performance of a combined cooling method of thermosyphons and insulation boards for tower foundation soils along the Qinghai–Tibet Power Transmission Line, Cold Regions Science and Technology, 121 (2016) 226-236. doi: https://doi.org/10.1016/j.coldregions.2015.06.006.
[18] Y. Qin, Y. Li, T. Bao, An experimental study of reflective shading devices for cooling roadbeds in permafrost regions, Solar Energy, 205 (2020) 135-141. doi: 10.1016/j.solener.2020.05.054.
[19] C. Beer, N. Zimov, J. Olofsson, P. Porada, S. Zimov, Protection of Permafrost Soils from Thawing by Increasing Herbivore Density, Scientific Reports, 10 (2020) 4170. doi: 10.1038/s41598-020-60938-y.
[20] I.M. Asanov, E.Y. Loktionov, Possible benefits from PV modules integration in railroad linear structures, Renewable Energy Focus, 25 (2018) 1-3. doi: 10.1016/j.ref.2018.02.003.
[21] E.Y. Loktionov, E.S. Sharaborova, I.M. Asanov, Prospective Sites for Solar-Powered Permafrost Stabilization Systems Integration in Russian Railways, 2019 8th International Conference on Renewable Energy Research and Applications (ICRERA), (2019) 8996544. doi: 10.1109/ICRERA47325.2019.8996544.
[22] T.-f. Hu, J.-k. Liu, Z.-h. Hao, J. Chang, Design and experimental study of a solar compression refrigeration apparatus (SCRA) for embankment engineering in permafrost regions, Transportation Geotechnics, 22 (2020) 100311. doi: 10.1016/j.trgeo.2019.100311.

[23] E.S. Sharaborova, T.V. Shepitko, E.Y. Loktionov, Experimental proof of a solar-powered heat pump system for permafrost thermal stabilization, Energies, (2021) under review. doi: 10.20944/preprints202112.0288.v1.
[24] H. Johra, K. Filonenko, P. Heiselberg, C. Veje, S. Dall'Olio, K. Engelbrecht, C. Bahl, Integration of a magnetocaloric heat pump in an energy flexible residential building, Renewable Energy, 136 (2019) 115-126. doi: 10.1016/j.renene.2018.12.102.
[25] U. Jakob, 6 - Solar cooling technologies, in: G. Stryi-Hipp (ed.) Renewable Heating and Cooling, Woodhead Publishing, 2016, pp. 119-136. doi: 10.1016/B978-1-78242-213-6.00006-0.
[26] O.S. Popel, S.E. Frid, S.V. Kiselyova, Y.G. Kolomiets, N.V. Lisitskaya, Climate data for renewable energy in Russia (Climatic database), MIPT Publishing, Moscow, 2010 (in Russian).
[27] Research... Research and applications reference book on the climate of the USSR, Hydrometeoizdat, Leningrad, 1989 (in Russian).
[28] V.M. Veselov, I.R. Priblyskaya, Specialized arrays for climate research, http://aisori.meteo.ru/ClimateR, accessed: 2020.11.18.
[29] V. Stolbovoi, I. McCallum, Land Resources of Russia, http://webarchive.iiasa.ac.at/Research/FOR/russia_cd/copyright.htm, accessed: 2020.11.18.
[30] T.A. Dauzhenka, I.A. Gishkeluk, Consistency of the Douglas – Rachford splitting algorithm for the sum of three nonlinear operators: application to the Stefan problem in permafrost soils, Applied and Computational Mathematics, 2 (2013) 100-108. doi: 10.11648/j.acm.20130204.11.
[31] A. Alekseev, G. Gribovskii, S. Vinogradova, Comparison of analytical solution of the semi-infinite problem of soil freezing with numerical solutions in various simulation software, IOP Conference Series: Materials Science and Engineering, 365 (2018) 042059. doi: 10.1088/1757-899x/365/4/042059.
[32] D.L. Powers, Boundary Value Problems and Partial Differential Equations, 6th ed., Academic Press, 2009.
[33] V.G. Kondratyev, N.A. Valiev, Stabilization of soil at central part of Baikal-Amur rairoad using solar and precipitation blinds, International conference Kazakhstan-Cold, 1-2/03/2016, (2016) 65-72.
[34] K. Solbakken, B. Babar, T. Boström, Correlation of wind and solar power in high-latitude arctic areas in Northern Norway and Svalbard, Renew. Energy Environ. Sustain., 1 (2016) 42. doi: 10.1051/rees/2016027.
[35] M. Angelopoulos, S. Westermann, P. Overduin, A. Faguet, V. Olenchenko, G. Grosse, M.N. Grigoriev, Heat and Salt Flow in Subsea Permafrost Modeled with CryoGRID2, Journal of Geophysical Research: Earth Surface, 124 (2019) 920-937. doi: 10.1029/2018jf004823.
[36] L. Jiang, Y. Song, L. Sun, C. Song, X. Wang, X. Ma, C. Liu, J. Gao, Effects of warming on carbon emission and microbial abundances across different soil depths of a peatland in the permafrost region under anaerobic condition, Applied Soil Ecology, 156 (2020) 103712. doi: 10.1016/j.apsoil.2020.103712.
[37] Z. Ci, F. Peng, X. Xue, X. Zhang, Permafrost Thaw Dominates Mercury Emission in Tibetan Thermokarst Ponds, Environmental Science & Technology, 54 (2020) 5456-5466. doi: 10.1021/acs.est.9b06712.
[38] K.R. Miner, J. D'Andrilli, R. Mackelprang, A. Edwards, M.J. Malaska, M.P. Waldrop, C.E. Miller, Emergent biogeochemical risks from Arctic permafrost degradation, Nature Climate Change, 11 (2021) 809-819. doi: 10.1038/s41558-021-01162-y.
[39] K.M. Walter, S.A. Zimov, J.P. Chanton, D. Verbyla, F.S. Chapin, Methane bubbling from Siberian thaw lakes as a positive feedback to climate warming, Nature, 443 (2006) 71-75. doi: 10.1038/nature05040.
[40] Y.A. Dvornikov, M.O. Leibman, A.V. Khomutov, A.I. Kizyakov, P. Semenov, I. Bussmann, E.M. Babkin, B. Heim, A. Portnov, E.A. Babkina, I.D. Streletskaya, A.A. Chetverova, A. Kozachek, H. Meyer, Gas-emission craters of the Yamal and Gydan peninsulas: A proposed mechanism for lake genesis and development of permafrost landscapes, Permafrost and Periglacial Processes, 30 (2019) 146-162. doi: 10.1002/ppp.2014.

[41] PVinsights, Solar PV Module Weekly Spot Price http://pvinsights.com/, accessed: 2020.09.17.
[42] Cost... Cost of thermal energy in Norilsk FY 2021, https://stroyfora.ru/tariff/area-d301a980-4a45-4066-9a20-4e783856a562/year-2021/type-9, accessed: 2021.12.21.